\documentclass[epj]{svjour}
\usepackage{epsf,graphicx,amssymb,amsmath}
\begin{document}
\title{A quasi-elastic regime for vibrated granular gases. }
\author{S. Auma\^\i tre\inst{1} \and A. Alastuey\inst{2} \and S. Fauve \inst{3}
}                     
%
%
\institute{Department of Physics and Astronomy, Haverford College, 370 Lancaster Avenue, Haverford, PA 19041 USA 
\and Laboratoire de Physique, 
CNRS UMR 5672, ENS, 46 all\'ee d'Italie,
69007 Lyon France 
\and Laboratoire de Physique Statistique, 
CNRS UMR 8550, ENS, 24 rue Lhomond,
75005 Paris France}
\date{Received: date / Revised version: date}
%
\abstract{
Using simple scaling arguments and two-dimensional numerical simulations of a granular gas excited by vibrating one of the container boundaries, we study a double limit of small $1-r$ and large $L$, where $r$ is the restitution coefficient and $L$ the size of the container. We show that if the particle density $n_0$ and $(1-r^2)(n_0 Ld)$ where $d$ is the particle diameter, are kept constant and small enough, the granular temperature, i.e. the mean value of the kinetic energy per particle, $\langle E \rangle/N$, tends to a constant whereas the mean dissipated power per particle, $\langle D \rangle/N$, decreases like $1/\sqrt{N}$ when $N$ increases, provided that $(1-r^2)(n_0 Ld)^2 < 1$.  The relative fluctuations of $E$, $D$ and the power injected by the moving boundary, $I$, have simple properties in that regime. In addition, the granular temperature can be determined from the fluctuations of the power $I(t)$ injected by the moving boundary.}

\PACS{
      {PACS-45.70.-n}{Granular systems}   \and
      {PACS-05.40.-a}{Fluctuation phenomena, random processes, noise, and Brownian motion}
     } 
%
\maketitle
\section{Introduction}
\label{intro}
Granular gases have been widely studied in the recent years as a canonical example of dissipative systems. Energy is lost at each collision between macroscopic particles but a statistically stationary regime can be reached with a continuous energy input, achieved in most experiments by vibrating the container boundary. In that regime, the power $I$ injected into the gas by the moving boundary, and the one dissipated by inelastic collisions, $D$, have equal time averages $\langle I \rangle = \langle D \rangle$ and the granular temperature, i.Êe. the mean kinetic energy per particle, $\langle E \rangle/N$ is constant in time. Several laboratory experiments \cite{exp} and numerical simulations \cite{num} have been performed to study the properties of the granular temperature. It has been shown that $I$ and $D$ display strong fluctuations \cite{Aumaitre01} and this has been confirmed experimentally \cite{Feitosa04}. Besides the obvious relation, $\langle I \rangle = \langle D \rangle$, there also exist relations between higher moments of the fluctuations of $I$ and $D$ and their large deviations \cite{Aumaitre04}. 

Like in most dissipative systems \cite{Aumaitre01}, the energy balance takes the form $\dot E = I(t) - D(t)$. Our aim here is to study a small dissipation limit such that $\langle I \rangle /N$ and  $\langle D \rangle /N$ vanish whereas $\langle E \rangle/N$ stays finite. 
We show that this can be achieved if the restitution coefficient, $r$, is scaled appropriately, $(1-r^2) (n_0 Ld) = \rm{constant}$ when the size $L$ of the container is increased at constant particle density $n_0$.  In that quasi-elastic limit, the granular temperature $\langle E \rangle/N$ becomes constant whereas the mean injected or dissipated power per particle scales like $1/\sqrt N$, provided that $(1-r^2) (n_0 Ld)^2 < 1$. The size of the system cannot become arbitrarily large, otherwise inhomogeneities appear as discussed in section 2. The probability density function (PDF) of the particle velocity becomes close to Gaussian and relative fluctuations of kinetic energy or injected power behave according to the law of large numbers. In addition, the temperature extracted from the fluctuations of injected power turns to be quite close to the granular temperature.

\section{Scaling arguments and the quasi-elastic limit}
\label{sec:1}

A two-dimensional granular gas is simulated with an event driven molecular dynamics method \cite{moldyn}. $N$ disks of mass $m$ and diameter $d$ are enclosed in a square box of size $L$. Energy input is provided by one vibrating wall with a symmetric saw-tooth motion, $y= \pm V_0t$, of period $T$.  Like in most previous studies, particle-wall collisions are elastic whereas inelastic binary collisions between particles are modeled with a constant restitution coefficient $r$, where $r$ is the ratio between pre- and post-collisional normal relative velocities. 

When $r$ is sufficiently close to one and the gas is sufficiently dilute, the medium is roughly homogeneous, $n(y) \approx n_0$. Then the mean dissipated and injected powers can be easily estimated. $\langle D \rangle$ can be related to the mean kinetic energy $\langle E \rangle$ by observing that an amount of energy $(1-r^2)\langle E \rangle/N$ is on average lost at each binary collision. This suggests $\langle D \rangle \propto (1-r^2) \langle E \rangle \nu_c$, where $\nu_c$ is the collision frequency per particle, $\nu_c \propto n_0 d v_{rms}$ whith $\langle E \rangle = mNv_{rms}^2 /2$.  The mean injected power $\langle I \rangle$ has been evaluated in several papers, with or without gravity \cite{Inj,McNamara97,num}.  For a symmetrical excitation, it has been shown that $\langle I \rangle \propto mV_0^2 \nu_P$, where $\nu_P$ is the collision frequency of the particles with the moving boundary \cite{Inj}. In the present conditions, $\nu_P \propto n_0 L v_{rms}$ and we obtain
$\langle I \rangle \propto mV_0^2 \,n_0 L v_{rms}$. Then, using $\langle I \rangle = \langle D \rangle$, we find for the granular temperature
\begin{equation}
\frac{\langle E \rangle}{N}  \propto \frac{mV_0^2}{(1-r^2)n_0 Ld}\,
\label{granulartemp}
\end{equation}
while the mean energy flux per particle is given by
\begin{equation}
\frac{\langle I \rangle}{N} = \frac{\langle D \rangle}{N} \propto  \frac{mV_0^3}{\sqrt{(1-r^2)dL}}\, \frac{1}{\sqrt N}.
\label{energyfluxperpart}
\end{equation}
We have shown that for $r=0.95$ and $r=0.9$, the above relations give correct predictions in the dilute limit \cite{Aumaitre06}. 
Notice that the elastic limit $r\rightarrow 1$ at fixed $L$ is singular since the granular temperature $E/N$ diverges. In order to obtain a finite limit, we must take the double limit  
$r \rightarrow 1$ and $N$ large with the constraint
\begin{equation}
(1-r^2)(n_0 Ld) = \rm{constant} = C \;\;  \rm{with} \;\; n_0 = \rm{constant}. 
\label{scaling}
\end{equation}
Then , the granular temperature is kept constant while the mean energy flux per particle from injection to dissipation decreases like $1/ \sqrt N$. 

Those simple behaviors no longer hold if the particle number and the size of the container become too large, the density being kept constant. Indeed, for $L \gg l_0$, a particle undergoes $(L/l_0)^2$ collisions when traveling on a distance $L$. Thus it dissipates an amount of energy of order $(n_0 Ld)^2 (1-r^2) \langle E \rangle /N $. The validity of the above scaling laws then require $(1-r^2) (n_0 Ld)^2 < 1$, otherwise the granular temperature and density profiles are non homogeneous. For a given value of $C$, this provides an upper bound for $L$ and $N$.

\begin{figure}
\resizebox{0.75\columnwidth}{!}{%
  \includegraphics{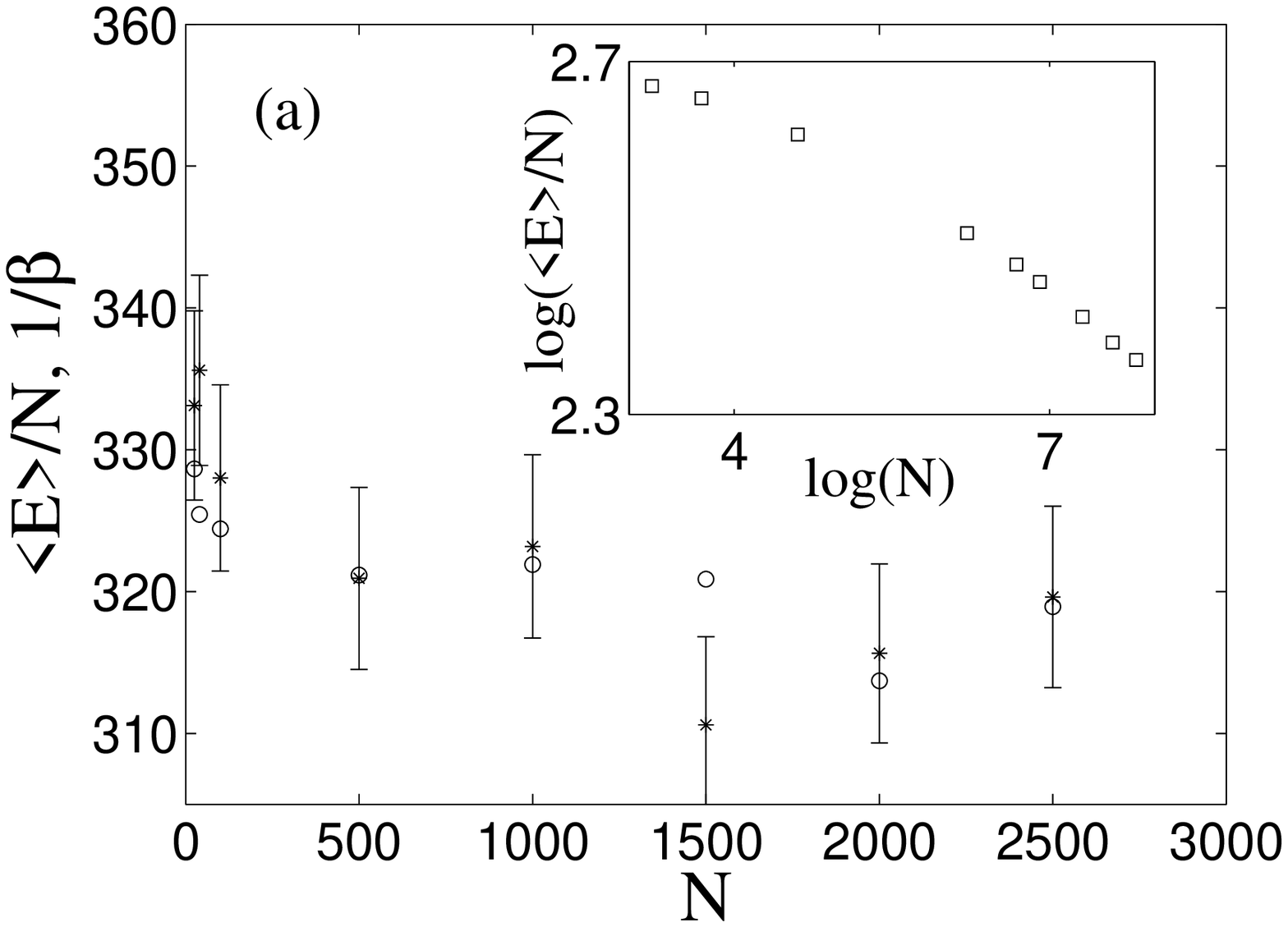}
}
\resizebox{0.75\columnwidth}{!}{%
  \includegraphics{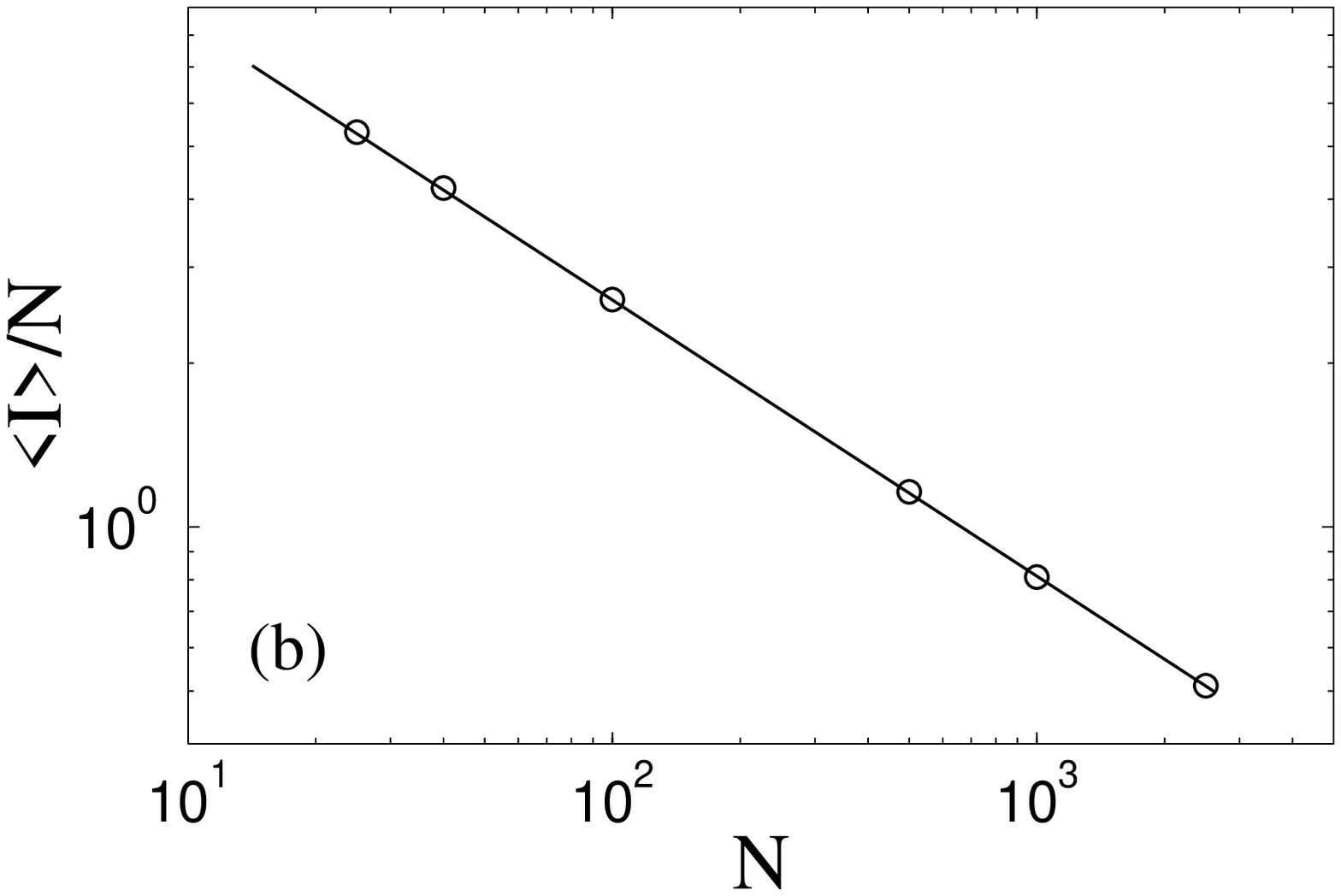}
}
\caption{a) Granular temperature $\langle E \rangle/N$  ($\circ$) and temperature $T_F= 1/\beta$  ($\ast$) extracted from the fluctuations of the injected power $I$ as a function of $N$ for fixed density, $n_0=N/L^2 = 0.01$ and $C= 0.025$. The inset shows that $\langle E \rangle/N$ does not reach a plateau if $n_0 =0.04$ and $C = 0.45$ b) Mean injected power per particle, $\langle I \rangle/N$ as a function of $N$, for fixed density, $n_0 = 0.01$ and $C= 0.025$. The full line has slope $-1/2$.}
\label{fig 1}       
\end{figure}

\section{Numerical results}
\label{sec:2}

We take $m$ and $T$ respectively as units of mass and time, $V_0=3$, and $d=0.5$. In most simulations, $n_0=0.01$ and $C = 0.025$. Then, we vary the particle number in the range $25 \leq N \leq 2500$. We observe in  Fig. 1a that the granular temperature is constant within $3\%$ in that range. Its numerical value ($320$) is in agreement with our order of magnitude estimate ($360$) from (\ref{granulartemp}). In addition, as shown in Fig. 1b, the energy flux per particle decreases according to $\langle I \rangle/N \propto 1/\sqrt N$ as expected. Thus, the above scaling laws can be observed on three decades in $N$ in the quasi-elastic limit (\ref{scaling}) whereas they are restricted to less than a decade when $N$ is changed with $n_0$ and $r$ kept constant \cite{Aumaitre06}. This results from the homogeneity of the system which is maintained for larger values of $N$ in the quasi-elastic limit (\ref{scaling}). In addition,  the PDF of the velocity fluctuations remains almost unchanged and close to Gaussian in that limit (see Fig. 2). The velocity $V$ along the vibration axis is less Gaussian than the one of the perpendicular component $U$ when $N$ is small. This slight anisotropy is not surprising and does not persist when $N$ increases. Although it is known that the PDF cannot be exactly Gaussian in general \cite{brey03}, it is very close to a Gaussian in the limit we are considering. That nearly equilibrium behavior is also displayed by relative fluctuations of energy $E$ and power $I$ or $D$. Their standard deviation $\sigma (X)$ related to their mean value are shown in Fig. 3. We observe that they are in agreement with the predictions of the law of large numbers, like for systems at equilibrium. Since $E$ and $D$ are quantities averaged on the whole surface $L^2$, their relative fluctuations decrease like $L^{-1}$ i.e. $1/\sqrt N$, whereas for $I$ averaged on the moving boundary, relative fluctuations decay like $N^{-1/4}$. Thus, we have $\sigma (E) \propto \sqrt N$, $\sigma (I) \propto N^{1/4}$ and $\sigma (D) = O(1)$. Those simple behaviors are not observed when $N$ is increased with $n_0$ and $r$ kept constant \cite{Aumaitre04}.

It is interesting to determine how the Gaussian behavior is lost far from the present limit. To wit, we take $C = 0.1$ instead of $0.025$. We obtain $\langle E \rangle /N$ roughly $4$ times smaller as expected from (\ref{granulartemp}) and the previous results remain qualitatively similar although higher moments of the velocity PDF's display a slightly less Gaussian behavior. We then take, $n_0 = 0.04$ and $C = 0.44$. In that case, $\langle E \rangle/N$ does not reach the plateau predicted by equation \ref{granulartemp} as shown in the inset of Fig. 1a. In addition, the PDF's of $V$ exhibits strongly non Gaussian behavior for the largest values of $N$ when $(1-r^2) (n_0 Ld)^2$ becomes of order $1$ (($\ast$) in Fig. 2). 

\begin{figure}
\resizebox{0.75\columnwidth}{!}{%
  \includegraphics{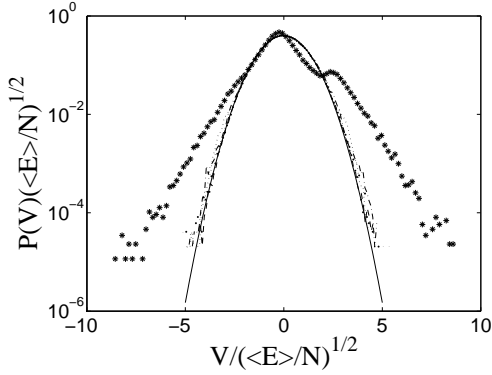}
}

\caption{PDF's of the velocity component $V$ parallel to the vibration axis, for $N= 25$ (dotted line), $N=40$ (dash-dotted line), $N=500$ (dots) and $N=2500$ (dashed line), with fixed density, $n_0 = 0.01$ and $C = 0.025$, $(1-r^2) (n_0 Ld)^2 \leq 0.0625$ and for $N = 2500$ ($\ast$) with $n_0 = 0.04$ and $C = 0.45$ thus $(1-r^2) (n_0 Ld)^2 = 2.25$. The full line is a Gaussian fit.}
\label{fig 2}       
\end{figure}

\begin{figure}
\resizebox{0.75\columnwidth}{!}{%
  \includegraphics{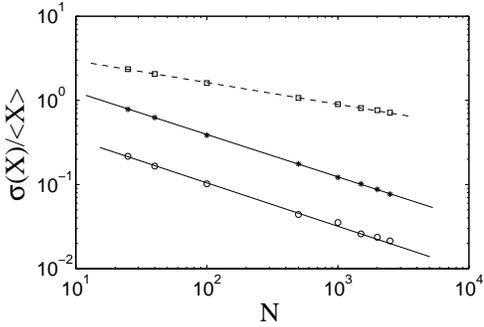}
}
\caption{Standard deviations of the injected power $I$ ($\square$), the dissipated power, $D$ ($\ast$), and the kinetic energy, $E$ ($\circ$), related to their respective mean values, as functions of $N$, for fixed density, $n_0 = 0.01$ and $C= 0.025$. The full lines have slope $-1/2$ while the dotted line has slope $-1/4$.}
\label{fig 3}       
\end{figure}

\section{Fluctuations of the injected power}
\label{sec:3}

Since the injected power $I(t)$ can be easily measured experimentally by recording the force applied by the particles on the moving boundary times its velocity, it has been proposed to try to extract some information on the system by studying its fluctuations \cite{Aumaitre01}. The PDF, $P(I)$, of the injected power $I(t)$ during one vibration period $T$ is displayed in Fig. 4 for different values of $N$.  It involves negative events corresponding to particles colliding with the moving boundary at velocity $-V_0$. The PDF's have a roughly Gaussian shape contrarily to previous ones obtained outside the present scaling range, which exhibit asymmetric exponential tails \cite{Aumaitre01,orsayboys}. This is specific to the limit considered here, where many uncorrelated collisions with the moving boundary occur during each period. Correspondingly, the standard deviation increases with $N$ like $\sigma(I) \propto N^{1/4}$.

We now consider the PDF, $P(I_{\tau} = \epsilon)$, of the injected power $I(t)$ averaged on a time interval $\tau$ for different values of $N$. In Fig. 5a, we draw centered PDF's reduced by the variance estimated as follows: according to large deviation analysis, for $\tau \gg \tau_I$ where $\tau_I$ is the correlation time of $I(t)$, $P$ is of the form, $P \propto \exp \left[\tau f(\epsilon) \right]$, where $f$ is maximum for $\epsilon = \langle I \rangle$.
Using the Gaussian approximation near $\langle I \rangle$, together with the previously observed scaling of the variance with $N$, we find the approximate form
\begin{equation}
P(I_{\tau} = \epsilon) \approx \frac{\sqrt{\tau}}{\sqrt{2\pi} \sigma_0 N^{1/4}} \, \exp \left[ - \frac{(\epsilon - \langle I \rangle)^2 \tau}{2 \sigma_0^2 \sqrt{N}} \right],
\label{Gaussapprox}
\end{equation}
valid in the neighborhood of $\langle I \rangle$.
In addition to scaling $\sigma^2(I_{\tau}) \propto \sqrt{N} / \tau$,  it is shown that the PDF's of $I_{\tau}$ for different values of $\tau$ and $N$ roughly follow a single curve when plotted as a function of the centered and reduced variable $(I_{\tau} - \langle I \rangle) \tau^{1/2} N^{-1/4}$ (see Fig. 5a).  However the whole distribution is skewed with a slightly larger positive tail, so it is not Gaussian.  

Fig. 5b shows that the quantity, $\frac{1}{\tau}\, \log {\left( P(\epsilon)/P(-\epsilon) \right)}$ is linear in $\epsilon$ up to values of $\epsilon$ larger than $\langle I \rangle$. Using (\ref{Gaussapprox}), we verify that the slope is $2 \langle I \rangle / \sigma_0^2 \sqrt{N}$, which does not depend on $\tau$ and $N$ (since $\langle I \rangle \propto \sqrt{N}$).
The inverse of the slope is homogeneous to a temperature $T_F = 1/ \beta$ which is plotted in Fig. 1a. We observe that $T_F \approx \langle E \rangle/N$ within $1\%$ in the range $500 \leq N \leq 2500$. This strongly suggest that the granular temperature can be extracted from the fluctuations of the injected power in the present limit. However, we emphasize that the fluctuation theorem \cite{FT} is not expected to hold for a dissipative granular gas \cite{Aumaitre01,orsayboys}. In general, $T_F \neq \langle E \rangle/N$ and their respective behaviors when $N$ is varied are also different \cite{Aumaitre04}. The present observation $T_F \approx \langle E \rangle/N$ is quite specific to the quasi-elastic limit (\ref {scaling}).

\begin{figure}
\resizebox{0.75\columnwidth}{!}{%
  \includegraphics{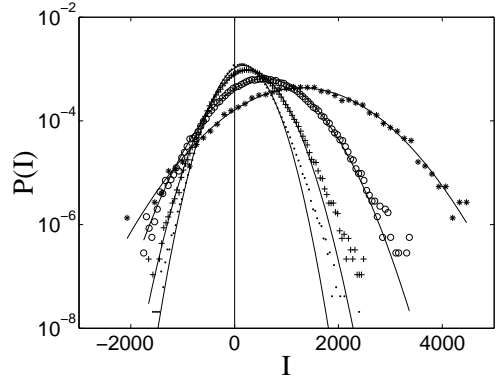}
}
\caption{PDF's of the injected power for $n_0 = 0.01$, $C= 0.025$ and different values of $N$:  $N=40$ ($\cdot$), $N=100$ (+), $N=500$ ($\circ$), $N=2500$ ($\ast$). The full lines are Gaussian fits.}
\label{fig 4}       
\end{figure}

\begin{figure}
\resizebox{0.75\columnwidth}{!}{%
  \includegraphics{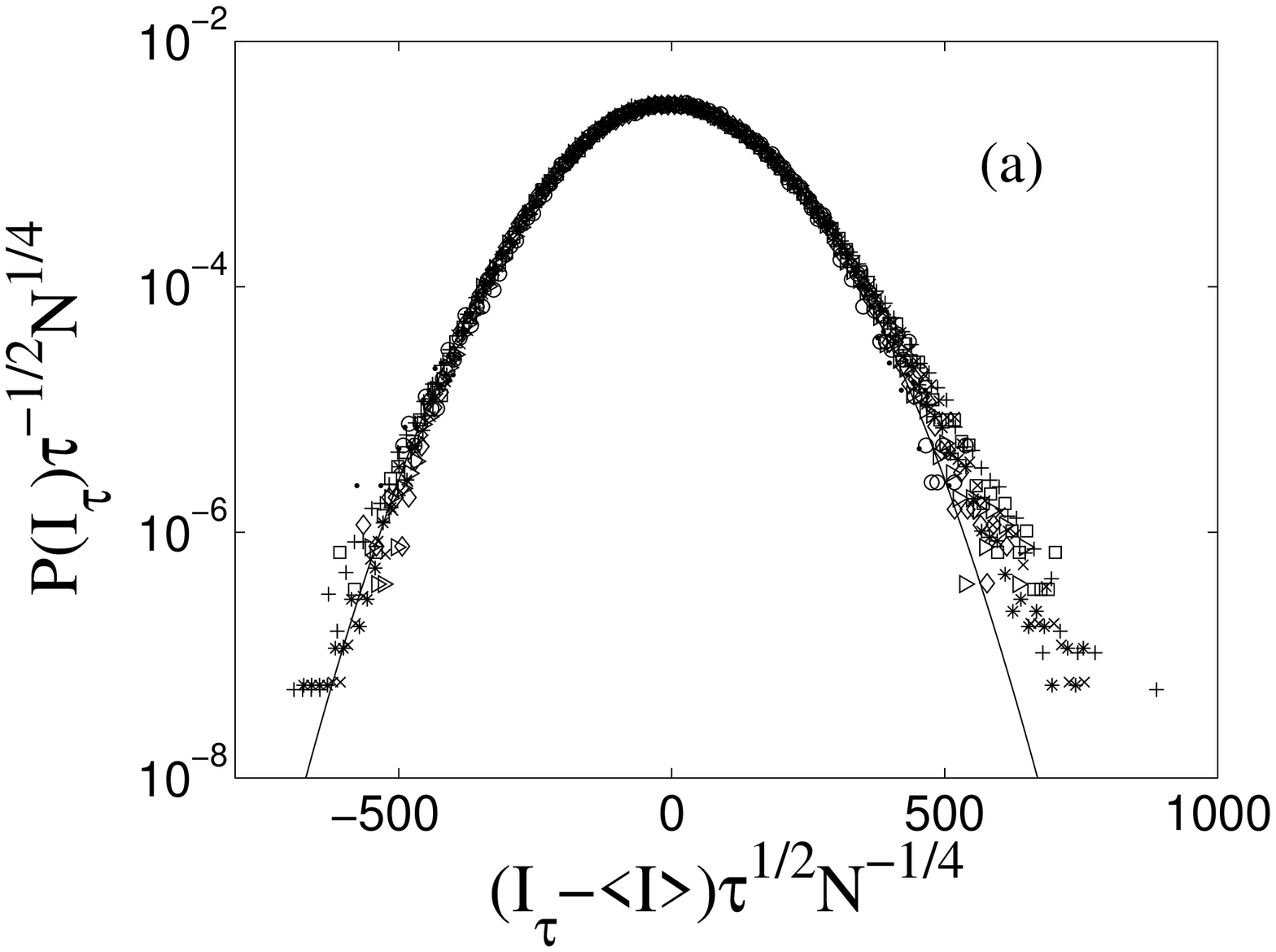}
}
\resizebox{0.75\columnwidth}{!}{%
  \includegraphics{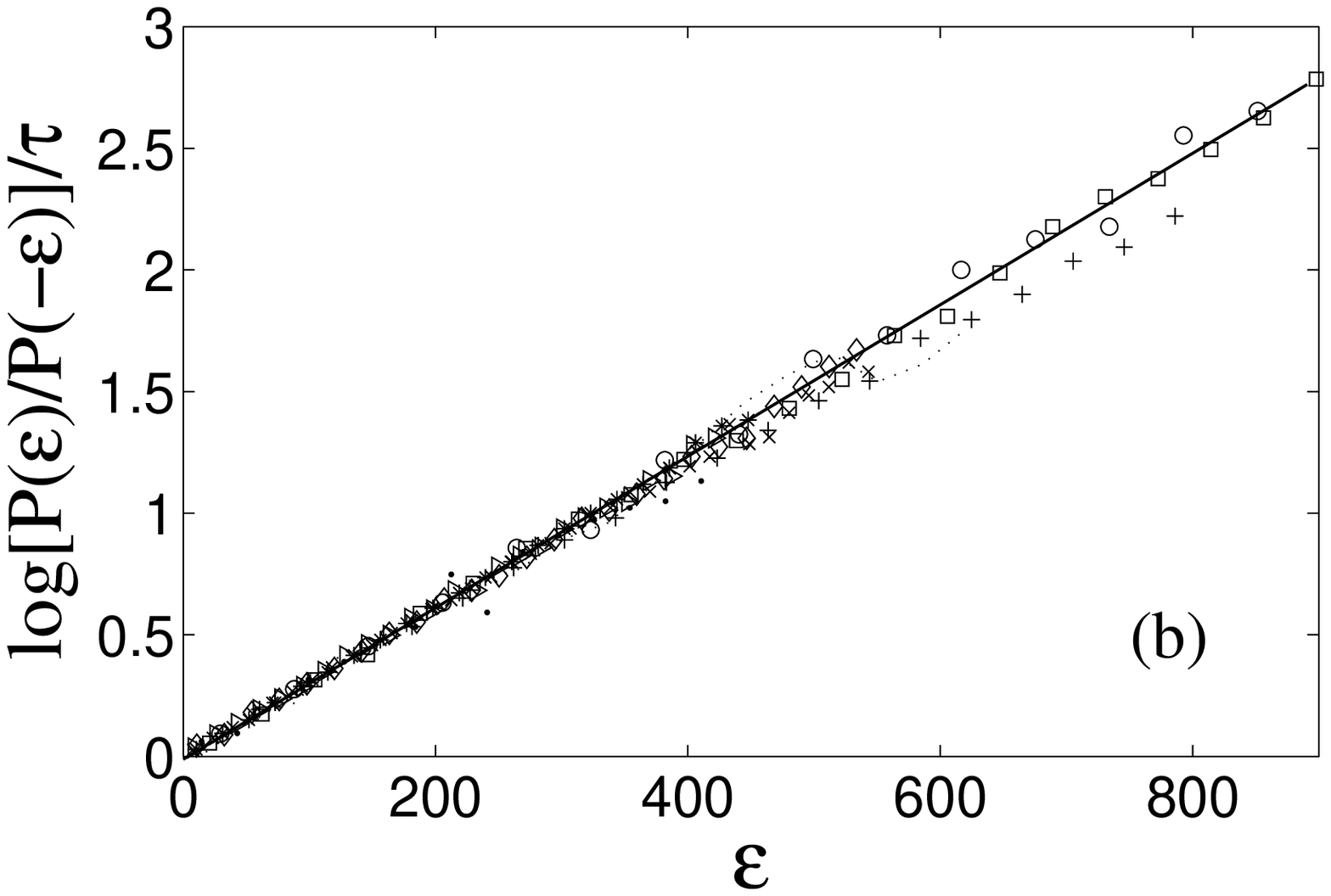}
}

\caption{a) PDF of $(I_{\tau} - \langle I \rangle) / (\sqrt{\sqrt{N} /\tau})$, for $n_0 = 0.01$, $C= 0.025$ and $N=40$, $\tau = T$ ($+$), $N=40$, $\tau = 3T$ ($\ast$), $N=40$, $\tau = 5T$ ($\times$), $N=100$, $\tau = T$ ($\square$), $N=100$, $\tau = 3T$ ($\lozenge$), $N=100$, $\tau = 5T$ ($\rhd$), $N=1000$, $\tau = T$ ($\circ$), $N=1000$, $\tau = 3T$ ($\dots$), $N=1000$, $\tau = 5T$ ($\bullet$). The full line is a Gaussian fit. 
 b) The quantity $\frac{1}{\tau}\,\log {\frac{P(I_{\tau} = \epsilon)} {P(I_{\tau} = -\epsilon)}}$, where $I_{\tau}$ is the injected power averaged on a time interval $\tau$,   is plotted as a function of $\epsilon$ (same symbols as for a).}
\label{fig 5}       
\end{figure}

\section{Discussion and concluding remarks}
\label{sec:4}

A double limit, involving $r \rightarrow 1$ together with $N \rightarrow \infty$, has been first considered in \cite{McNamara92} for a one-dimensional granular medium and called the quasi-elastic limit. It has been shown that the leading order correction to the Boltzmann equation for particles without interaction is of order $(1-r) N$, so in the limit $N \rightarrow \infty$ with $N(1-r)$ kept constant and small enough, inelastic collapse of particles is avoided. The same condition for the absence of inelastic collapse has been obtained in higher space dimension $D$, $(1-r) n_0 L d^{D-1}$ small enough \cite{Esipov97}. The scaling used here $(1-r^2) n_0 L d^{D-1}$ = constant, is only slightly different. However, it has been derived from the energy balance in an excited granular gas instead of considering the Boltzmann equation for a freely cooling gas. We also mention that a more restrictive condition, $(1-r) \left(n_0 L d^{D-1}\right)^2$ small enough has been given in \cite{Esipov97} for a gas like system, i.e. non condensed. A condition similar to the latter can be understood in a simple way within an hydrodynamic limit. It can be also written $(1-r^2) \ll K^2$ where $K=l_0 /L$ is the Knudsen number. This is sometimes considered as a requirement for the validity of the hydrodynamic description of granular media \cite{Goldhirsch03}, although this is still a matter of debate \cite{Brey98}. 

Our simulations show that the constraint $(1-r^2) n_0 L d^{D-1}$ = constant in the dilute limit, allows us to observe a fairly large range of $N$ for which the velocity PDF are roughly Gaussian although the forcing remains purely deterministic (no thermostat or applied fluctuating force). In that limit, the granular temperature remains constant as the size of the container is increased at constant density. The mean dissipated power per particle decreases like $1/\sqrt N$ and the relative fluctuations of energy or power behave like in systems at equilibrium. In addition, an out of equilibrium temperature, defined from the fluctuations of the injected power, becomes equal to the granular temperature. Those simple behaviors no longer hold if the particle number and the size of the container become too large, the density being kept constant. One must also keep $(1-r^2) \left(n_0 L d^{D-1}\right)^2 < 1$. When that parameter becomes of order one, the velocity PDF is non Gaussian, the fluctuations do not scale any more according to the law of large numbers and the temperature extracted from the fluctuations of injected power becomes different from the granular temperature. In the quasielastic equilibrium-like regime, it would be interesting to determine if the non-equipartition of energy observed in binary granular systems still holds \cite{binary}, or on the contrary, if other typical equilibrium properties are observed.

\end{document}